\documentclass[twocolumn,]{aastex62}

\hypersetup{linkcolor=red,citecolor=blue,filecolor=cyan,urlcolor=gray}
\usepackage{graphicx,amsmath,amssymb,amstext,natbib}
\usepackage{epsf}
\usepackage{epstopdf}
\graphicspath{{./}{figures/}}

\submitjournal{ApJ}

\shorttitle{ALMA 3\,mm number counts}
\shortauthors{Jorge A. Zavala et al.}
\begin{document}

\title{\sc Constraining the volume density of Dusty Star-Forming Galaxies through\\ 
 the first 3\,mm Number Counts from ALMA}

\correspondingauthor{Jorge A. Zavala}
\email{jzavala@utexas.edu}

\author[0000-0002-7051-1100]{J. A. Zavala}
\affil{The University of Texas at Austin, 2515 Speedway Blvd Stop C1400, Austin, TX 78712, USA}

\author[0000-0002-0930-6466]{C. M. Casey}
\affil{The University of Texas at Austin, 2515 Speedway Blvd Stop C1400, Austin, TX 78712, USA}

\author[0000-0001-9759-4797]{E. da Cunha}
\affil{ Research School of Astronomy and Astrophysics, The Australian National University, Canberra ACT 2611, Australia}

\author[0000-0003-3256-5615]{J. Spilker}
\affil{The University of Texas at Austin, 2515 Speedway Blvd Stop C1400, Austin, TX 78712, USA}

\author[0000-0002-8437-0433]{J. Staguhn}
\affil{NASA Goddard Space Flight Center, Code 665, Greenbelt, MD 20771, USA}
\affil{Bloomberg Center for Physics and Astronomy, Johns Hopkins University 3400 N. Charles Street, Baltimore, MD 21218, USA}

\author[0000-0001-6586-8845]{J. Hodge}
\affil{Leiden Observatory, Niels Bohrweg 2, 2333 CA Leiden, The Netherlands}

\author[0000-0003-3627-7485]{P. M. Drew}
\affil{The University of Texas at Austin, 2515 Speedway Blvd Stop C1400, Austin, TX 78712, USA}



\begin{abstract}
We carry out a blind search of 3\,mm continuum sources using the ALMA Science Archive to derive the first galaxy number counts at this wavelength. The analyzed data are drawn from observations towards three extragalactic legacy fields: COSMOS, CDF-S, and the UDS comprising more than 130 individual ALMA Band 3 pointings and an effective survey area of $\approx200\rm\,arcmin^2$ with a continuum sensitivity that allows for the direct detection of unlensed Dusty Star-Forming Galaxies (DSFGs) dust emission beyond the epoch of reionization. We present a catalog of 16 sources detected at $>5\sigma$ with flux densities $S_{\rm 3mm}\approx60-600\rm\,\mu Jy$ from which  number counts are derived. 
These number counts are then used to place constraints on the volume density of DSFGs with an empirical backward evolution model. Our measured 3\,mm number counts indicate that the contribution of DSFGs to the cosmic star formation rate density at $z\gtrsim4$ is non-negligible. This is contrary to the generally adopted assumption of a sharply decreasing contribution of obscured galaxies at $z>4$ as inferred by optical and near-infrared surveys. This work demonstrates the power of ALMA 3\,mm  observations  which can reach outstanding continuum sensitivities during typical spectral line science programs. Further constraints on 3\,mm-selected galaxies will be essential to refine models of galaxy formation and evolution as well as models of early Universe dust production mechanisms.
\end{abstract}


\section{Introduction} \label{sec:intro}

Understanding the star formation activity across cosmic time is among the most important goals of modern observational and theoretical astrophysics.
 
Since around half of  optical and UV stellar radiation in galaxies is absorbed by dust and re-emitted at 
far-infrared (IR) and (sub-)millimeter wavelengths,
the achievement of a complete unbiased census of the Universe's star formation activity requires a multi-wavelength approach that reconciles both  obscured and unobscured pictures of
the Universe. While the mapping of cosmic star formation was forged on stellar emission, (sub-)millimeter surveys (beginning with \citealt{Smail1997a,Barger1998a,Hughes1998a}) have shown us that the majority of the star formation 
activity at its peak epoch is primarily enshrouded by dust (\citealt{Madau2014a}). 
 However,  while studies of galaxies' rest-frame UV/Optical emission span out to $z\sim11$ (e.g. \citealt{Ellis2013a,Oesch2013a,Bouwens2015a,Finkelstein2015a,Finkelstein2016a}),  
 our knowledge of the prevalence of dust-obscured star formation at these 
earlier epochs is completely unconstrained
due to the lack of complete samples of $z\gtrsim4$ dusty star-forming galaxies (DSFGs, see review by \citealt*{Casey2014a}).

Though the current large area (sub-)millimeter surveys (like those carried out by {\it Herschel} Space Observatory and the South Pole Telescope, SPT) have made surprising discoveries of DSFGs up to $z\approx6-7$ (\citealt{Riechers2013a,Strandet2017a,Zavala2018a}), they are only sensitive to the rarest, most extreme  starbursts with star formation rates (SFRs) of  $\gtrsim1000\,\rm M_\odot\,yr^{-1}$,  or to gravitationally amplified galaxies whose volume density is difficult to constrain. Less extreme galaxies with SFRs of hundreds of solar masses per year, can in principle be detected in deeper (but smaller area) maps already in-hand from single-dish (sub-)millimeter telescopes (e.g. \citealt{Geach2017a}).  Nevertheless, the DSFGs identified by these observations (at typical wavelengths of $\lambda=850\,\mu\rm m - 1.1\,mm$) are overwhelmed  by the abundant  population of $1<z<3$ DSFGs (e.g. \citealt{Michaowski2017a,Zavala2018b}), making the identification of the most distant objects a very challenging task (not to mention the large positional uncertainties of single-dish telescopes). The deep pencil-beam surveys from ALMA (e.g. \citealt{Umehata2015a,Aravena2016a,Hatsukade2016a, Walter2016a,Dunlop2017a,Franco2018a,Hatsukade2018a}) are also dominated by low-redshift sources because of the small survey area and selection wavelength (see discussion by \citealt{Casey2018b}). 
As a consequence, our knowledge of the physical properties and the space density of more moderate luminosity DSFGs with $100\lesssim\rm SFRs\lesssim1000\rm\,M_\odot\,yr^{-1}$ at high redshifts, and consequently their contribution to the cosmic star formation rate density (CSFRD), is still unknown. Alternative strategies are therefore necessary to characterize the population of DSFGs at the highest redshifts. 
This is of high importance not only to derive a complete census of the CSFRD but also to  shed light 
on early Universe dust production mechanisms and the origin of the Universe's first massive galaxies.

The combination of model predictions and integrated measurements  such as the number counts,
can be used to derive robust constraints on the space density of a population of galaxies (e.g. \citealt{Bethermin2012a,Bethermin2017a, Hayward2013a,Cowley2015a}), even when individual redshifts of galaxies are not available. Our recently developed empirically motivated backward evolution model of the (sub-)millimeter sky (\citealt{Casey2018a,Casey2018b}) adopts an evolving infrared galaxy luminosity function (IRLF) between $0<z\lesssim10$ to make predictions, as a function of (sub-)millimeter wavelength and depth, of the number counts and redshift distribution of galaxies selected in the far-infrared (FIR) and (sub-)millimeter regime. As thoroughly discussed in \citealt{Casey2018a,Casey2018b}, the constraints provided by all the current submillimeter and millimeter surveys from both single-dish and interferometric observations are not tight enough to draw strong conclusions on the shape of the IRLF (and hence on the contribution of these galaxies to the CSFRD) at $z>2.5$. This lack of constraining power is illustrated by the fact that the aggregate of two decades of data at (sub-)millimeter wavelengths cannot distinguish between two extreme hypothetical scenarios: a dust-rich Universe where the CSFRD at $z>4$ is dominated ($\gtrsim90\,$\%) by DSFGs and an alternate dust-poor early Universe where dust-obscured star formation at $z>4$ is negligible (see \citealt{Casey2018a,Casey2018b}).
An important corollary of these studies suggests that surveys at longer wavelengths than those carried out in the past, specifically observations at 2\,mm  (e.g. \citealt{Staguhn2014a}) and 3\,mm, represent a promising way to identify and characterize the high-redshift population of DSFGs by effectively filtering out low-redshift sources.

This work represents one of the first efforts to exploit 3\,mm continuum observations for the detection of such distant objects. Selection at 3\,mm is an extension of the submillimeter-galaxy selection technique to the extreme. Indeed, galaxies found at 3\,mm are unlikely to lie at $z < 2$ due to the very strong millimeter negative {\it K}-correction (Figure \ref{fig:area_vs_depth}, see also \citealt{Casey2018b}). Though the detection of these galaxies requires very deep observations (since 3\,mm flux density arises from the faint tail of the Rayleigh-Jeans regime of the dust thermal emission), this depth is routinely achieved in ALMA spectroscopic surveys upon collapsing data cubes across the spectral dimension. Indeed, five 3\,mm-selected continuum sources have already been reported in the recent literature: one in the ASPECS-Pilot survey with a redshift of $2.543$ (\citealt{Aravena2016a,Walter2016a}) and four revealed conducting a spectral program analyzing source multiplicity in DSFGs (\citealt{Wardlow2018a}).
Here we report the results of a blind search for 3\,mm-detected sources, as discussed in \S\ref{sec:data_analysis},  and the first estimation of the 3\,mm galaxy number counts derived from ALMA observations, which is presented in \S\ref{sec:num_counts}. These sources were found in ALMA archival datasets covering a total solid angle of $198\,\rm arcmin^2$ in three different extragalactic survey fields: UDS, CDF-S, and COSMOS. The constraints provided by the number counts on the IRLF are described in \S\ref{sec:models_irlf} as well as the estimated dust obscured star formation rate density. Finally, our conclusions are presented in \S\ref{sec:conclusions}.

We assume a Planck cosmology throughout this paper, 
adopting $H_0 = 67.7\rm\,km\,s^{-1}\,Mpc^{-1}$ and $\Omega_\Lambda= 0.69$ 
(\citealt{Planck-Collaboration2016a}), and the \citet{Chabrier2003a} initial mass function (IMF) for SFR estimations.

\section{Data retrieval, analysis, and characterization} \label{sec:data_analysis}
Galaxies' 3\,mm dust continuum emission is expected to be several times fainter than their flux densities measured at shorter wavelengths (like the standard (sub-)millimeter wavebands at $\lambda\approx850-1100\rm\,\mu m$) due to the shape of the dust spectral energy distribution (SED) on the Rayleigh-Jeans tail. For example, a typical galaxy in the \citet{Casey2018a,Casey2018b} simulations at $z\sim2$ has a flux ratio of $S_{\rm1.2mm}/S_{\rm3mm}\sim25$, or a flux ratio of $S_{\rm1.2mm}/S_{\rm3mm}\sim10$ at $z\sim5$.  For this reason, 3\,mm observations are not an efficient method for detecting dust continuum emitters blindly, and therefore, no 3\,mm continuum-only blank field exists to date. However, most of the  spectroscopic studies of  molecular gas in low and high redshift galaxies are conducted in this waveband due to the large coverage of $^{12}\rm CO$ transitions (see, for example, figure 1 from \citealt{Walter2016a}). Since these spectral observations require deep sensitivity across relatively narrow frequency channels (with a velocity resolution on the order of $\approx10-100\rm\,km\,s^{-1}$), the typical achieved continuum depth across the total 8\,GHz bandwidth, in the case of ALMA, is enough to detect galaxies' dust emission up to very high redshifts, as shown in Figure \ref{fig:area_vs_depth}.
This work exploits ALMA archival Band 3 data to perfom a blind search of 3\,mm continuum-detected galaxies to derive the first number counts and to constrain the volume density of DSFGs.

\subsection{ALMA Band 3 archival data}\label{sec:ALMAdata}
\begin{figure}[h]
\includegraphics[width=0.45\textwidth]{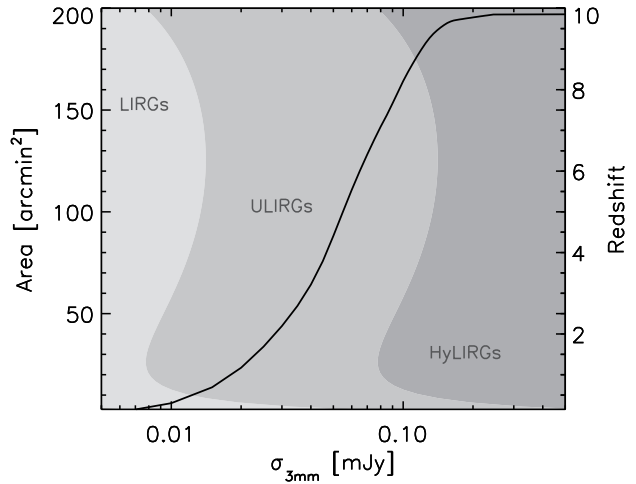}
\caption{The cumulative area covered by our survey as a function of $1\sigma$ r.m.s. depth is represented by the black solid line. Additionally, the corresponding luminosity detection limit at $5\sigma$ is shown as a function of redshift assuming a typical DSFG SED (a gray body with $T_D=35\,$K and $\beta=1.8$), including the impact of CMB (\citealt{da-Cunha2013a}). Three luminosity ranges are illustrated: LIRGs ($ 10^{11}\le\rm L_{\rm IR}< 10^{12}\rm\,L_\odot$), ULIRGs ($ 10^{12}\le\rm L_{\rm IR}< 10^{13}\rm\,L_\odot$), and HyLIRGs ($\rm L_{\rm IR}\ge 10^{13}\rm\,L_\odot$).
Given the strong negative {\it K}-correction in the 3\,mm band, we are sensitive to galaxies galaxies with $\rm L_{\rm IR}\gtrsim 10^{12} - 10^{13} \rm\,L_\odot$ up to $z\sim10$. \label{fig:area_vs_depth}}
\end{figure}

Using the ALMA Science Archive Query, we search for public ALMA Band 3 observations, which cover a frequency range of $\nu=84-116$\,GHz or $\lambda=2.59-3.57$\,mm.
We focused only on data acquired from Cycle 3 onwards, when the ALMA Science Pipeline was already commissioned, and  continuum maps were also processed and available through the archive. 
% Otherwise, the process of downloading, calibrating, and imaging the raw data would be very time consuming and would require a huge volume of computational resources.
To avoid contamination from Galactic sources, the search was limited to programs carried out within three well-known cosmological fields: UDS, CDF-S, and COSMOS, which comprise the vast majority of extragalactic Band 3 science pointings. Further, these fields have  exquisite ancillary multi-wavelength data that allows a detailed characterization of the detected 3\,mm sources (e.g. \citealt{Laigle2016a}). A restriction on the angular resolution of the images, $\theta\ge1.0$\,arcsec, was also imposed in order to avoid the incompleteness effects associated with high angular resolutions (see \citealt{Franco2018a}) and to avoid resolving out the emission of the galaxies. Finally, observations were restricted to have a continuum sensitivity of $\sigma_{\rm rms}<0.2\rm\,mJy\,beam^{-1}$, which roughly corresponds to the minimum depth required for the detection of unlensed DSFGs with SFRs$\lesssim1000\rm\,M_\odot\,yr^{-1}$ (see Figure \ref{fig:area_vs_depth}). 

After removing spatially-overlapping observations and projects with no continuum images available, a total of 135 maps were retrieved from almost 20 different projects (up to a public release date of May 2018), including not only single pointings but also mosaics made of several pointings (all ALMA project codes are reported in the Acknowledgments Section). This compilation covers an effective area  of $198\,\rm arcmin^2$, equivalent to the area encompassed by $\sim240$ ALMA Band 3 pointings within the primary beam FWHM ($\theta_{\rm FWHM}\approx60''$). This is an order of magnitude larger than the typical contiguous blank fields achieved with this facility (e.g. \citealt{Umehata2015a,Hatsukade2016a, Walter2016a,Dunlop2017a,Hatsukade2018a}). Figure \ref{fig:area_vs_depth} shows the total area analyzed in this work as a function of depth\footnote{The quoted depth of the observations and the flux densities of the detected sources have been scaled to 3\,mm (99.9\,GHz, the central frequency of Band 3) assuming $S_\nu\propto\nu^{2+\beta}$, with $\beta=1.8$ (a modified Rayleigh-Jeans law). This correction is usually of the same order (or less) than the typical flux boosting factor and/or the typical flux uncertainty. \label{footnote1}}. 

The primary science goal of most of these projects was to detect spectroscopic features, particularly CO emission lines, in targets selected from heterogeneous criteria (e.g. optically-selected galaxies, blank-fields, proto-cluster and cluster environments, etc.). This sample selection does not introduce any obvious bias in our blind search, since we are targeting continuum-selected galaxies which are expected to be high-redshift ($z>2$) DSFGs (see \S\ref{sec:source_extraction} and Appendix \ref{appendix}). In fact, as revealed by a quick visual inspection, only a few of the original targets are detected in the continuum images. These sources are not included in our source catalog.

Since continuum observations are not the primary science goal of the original projects, a further test on the quality of the retrieved continuum images was conducted. For maps where continuum sources were detected (see \S\ref{sec:source_extraction}), we individually re-reduce the raw data using {\sc casa}  (\citealt{McMullin2007a}) following the standard procedure, with a natural weighting of the visibilites in order to  maximize the sensitivity to faint sources.  The measured flux densities of the detected sources (see Table \ref{table:catalogue}) are in very good agreement with the values measured on the maps available through the archive  - although the signal-to-noise ratio (SNR) is typically lower in the retrieved images since a Briggs weighting is usually adopted.

\subsection{Source extraction and source catalog}\label{sec:source_extraction}

Source extraction was performed using the uncorrected primary beam continuum maps (which have the benefit of a constant noise) within a radius of $\approx1.3$ times the FWHM of the primary beam, where the antenna response sensitivity is $\ge0.3$. A central square mask with a side's dimension of 2 times the size of the synthesized beam is also applied on the primary target of each program. 

To search for source candidates, 
the uncorrected primary beam map of each observation is first divided by the noise of the same image to obtain a signal-to-noise map. The noise is assumed to be the 
$68^{th}$ percentile of the distribution of pixel values of the map, which corresponds to $1\sigma$ for a Gaussian distribution\footnote{The noise distribution of ALMA observations has been shown to be well described by Gaussian noise, especially in the
case of unresolved or marginally-resolved faint sources; e.g. \citealt{Dunlop2017a}.}. The $68^{th}$ percentile is preferred over the standard deviation of the map  since the latter can be overestimated by the presence of real sources. Then, source candidates are identified by searching pixels above a signal-to-noise threshold,  and the associated flux density of each candidate is measured  from the primary-beam corrected image at the same position and its error is assumed to be the noise 
measured in the whole uncorrected primary beam map divided by the primary beam response at the same pixel. Although values of $\rm SNR\approx3-4$ have been used in the literature (e.g. \citealt{Hodge2013a,Fujimoto2016a}), the large number of independent beams in ALMA maps, compared to single-dish observations, produces a significant contamination rate at these low SNRs (\citealt{Dunlop2017a}). Thus, here we adopt a conservative threshold of $5\sigma$ to minimize the contamination fraction (see \S\ref{sec:simulations}). Finally, a mask of 2 times the FWHM of the synthesized beam is applied at the position of the source candidate before repeating the process again until no more $>5\sigma$ pixels are found.

The 16 serendipitously-detected 3\,mm sources at $>5\sigma$ are reported in Table \ref{table:catalogue} along with the their individual SNRs, flux densities, and their associated uncertainties. This catalog includes the previously detected source in the ASPECS survey (\citealt{Walter2016a}) and three of the sources found in \citet{Wardlow2018a}. The remaining detection reported by \citeauthor{Wardlow2018a}, ALESS 49.C, falls just below our adopted threshold and hence is not included in the catalog.

\begin{deluxetable*}{lcccccr}[t]
\tablecaption{3\,mm ALMA Archival Survey source catalog. \label{table:catalogue}}
\tablecolumns{6}
\tablenum{1}
\tablewidth{0pt}
\tablehead{
\colhead{ID} &
\colhead{RA} &
\colhead{Dec} &
\colhead{SNR} &
\colhead{$S_{\rm 3mm}$\tablenotemark{a}} & \colhead{$z_{\rm spec}$} & \colhead{Other names}  \\
\colhead{} & \colhead{[hh:mm:ss.s]} & \colhead{[$^\circ:\arcmin:\arcsec$]} & \colhead{} & \colhead{[$\rm\mu Jy$]} & \colhead{} & \colhead{}
}
\startdata
ALMA-3mm.01\tablenotemark{$b$} & 03:31:09.8 & $-$27:52:25.6 &   24.0 & $240\pm10$ &  -                         &  ALESS\,41.C\tablenotemark{$c$}         \\
ALMA-3mm.02$ $                 & 02:16:44.3 & $-$05:02:59.7 &   8.7  & $118\pm14$ & -                          &   \\
ALMA-3mm.03$ $                 & 03:32:38.6 & $-$27:46:34.5 &   8.5  & $57\pm7$   &   2.54\tablenotemark{$d$}  &  ASPECS-3mm.1\tablenotemark{$d$}        \\
ALMA-3mm.04$ $                 & 02:17:42.8 & $-$03:45:31.2 &   7.4  & $130\pm18$ &  -                         &           \\
ALMA-3mm.05\tablenotemark{$e$} & 10:01:30.7 & $+$02:18:41.4 &   6.8  & $129\pm19$ & -                          &           \\
ALMA-3mm.06$ $                 & 03:31:02.9 & $-$28:42:29.8 &   6.7  & $117\pm17$ &  -                         &         \\
ALMA-3mm.07$ $                 & 03:31:26.7 & $-$27:56:01.0 &   6.6  & $53\pm8$   &  -                         & ALESS\,75.C\tablenotemark{$c$}          \\
ALMA-3mm.08$ $                 & 10:00:54.5 & $+$02:34:36.2 &   6.2  & $164\pm26$ &  4.55\tablenotemark{$f$}   & AzTEC-C17\tablenotemark{$g$}         \\
ALMA-3mm.09$ $                 & 03:32:50.7 & $-$27:31:34.7 &   6.1  & $63\pm10$  &  -                         &  ALESS\,87.C\tablenotemark{$c$}         \\
ALMA-3mm.10$ $                 & 02:16:44.5 & $-$05:02:21.6 &   5.9  & $91\pm16$  &   -    &   S2CLS-UDS.0074\tablenotemark{$h$}, ASXDF1100.003.1\tablenotemark{$i$}   \\
ALMA-3mm.11\tablenotemark{$e$} & 10:00:33.3 & $+$02:26:01.2 &   5.4  & $126\pm23$ &  2.51\tablenotemark{$j$}   &  AzTEC-C80b\tablenotemark{$k$}         \\
ALMA-3mm.12$ $                 & 10:00:34.4 & $+$02:21:21.7 &   5.4  & $125\pm23$ & 2.99\tablenotemark{$j$}    &           \\
ALMA-3mm.13$ $                 & 03:30:56.0 & $-$28:43:04.1 &   5.3  & $104\pm19$ &  -                         &          \\
ALMA-3mm.14$ $                 & 03:32:49.5 & $-$27:32:07.6 &   5.2  & $98\pm19$  &  -                         &           \\
ALMA-3mm.15$ $                 & 10:00:22.4 & $+$02:31:38.7 &   5.2  & $610\pm120$&   -                        &          \\
ALMA-3mm.16$ $                 & 10:02:00.1 & $+$02:24:18.1 &   5.0  & $263\pm52$ &  -                         &           
\enddata
% \tablenotetext{a}{Measured flux density scaled at 3\,mm (see Footnote\ref{footnote1}).}
% \tablenotetext{b}{Flux densities have be.}\tablenotetext{c}{Flux densities have be.}
\tablecomments{$^a$Measured flux density scaled to 3\,mm (see Footnote \ref{footnote1}).
$^b$This source was not included in the number counts due to the non-thermal emission (see \S\ref{sec:source_extraction}). 
$^c$\citet{Wardlow2018a}.
$^d$\citet{Walter2016a}.
$^e$This source was not included in the number counts since it may be physically associated with the primary target of the observations.
$^f$\citet{Schinnerer2008a}.
$^g$\citet{Aretxaga2011a}.
$^h$ALMA project code: 2015.1.01528.S.
$^i$\citet{Ikarashi2017a}.
$^j$Zavala et al. in preparation.
$^k$\citet{Brisbin2017a}.  
}
\end{deluxetable*}

A thorough investigation of the potential contaminants is important to ensure the purity of the catalog. 
As discussed by \citet{Wardlow2018a}, it is possible that sources' 3\,mm emission arises from non-thermal processes. Actually, ALMA-3mm.01 (also known as  ALESS\,41.C) shows an SED consistent with a flat-spectrum radio quasar and might be associated with a known radio source. This object has therefore not been included in the number counts estimation described below.  To rule out the possibility of including any other source with a non-thermal SED we re-reduce the ALMA data and create two continuum maps for each source with the spectral windows corresponding to the low and high frequency sidebands, respectively. All of the sources show properties consistent with thermal emission (i.e. $S_{\nu_{\rm high}}/S_{\nu_{\rm low}}>1$), with possible exceptions of ALMA-3mm.15 and ALMA-3mm.16, for which the low SNRs on the individual (split) continuum maps prevent a robust determination; their colors might be consistent with a flat-spectrum (although at low $\lesssim3\sigma$ significance).
A further analysis of the extracted spectrum for these objects revealed that sources ALMA-3mm.05 and ALMA-3mm.11 may be physically associated with, or at the same redshifts as, the original targets of their observations, based on presence of millimeter emission lines at similar frequencies (Zavala et al. in preparation). These two sources are hence not considered in the number counts estimation aimed at reporting an unbiased statistic, but are reported in the catalog given the adopted selection criterion. 
A deeper analysis of the nature of these sources as well as a full characterization of their physical properties require a multi-wavelength analysis which will follow in a future paper. In the meantime, a brief description of the 13 sources used in the number counts analysis is presented in Appendix \ref{appendix}, emphasizing the potential biases introduced by the original targets of these observations which might affect the {\it blindness} of this survey.

\subsection{Completeness, flux boosting, and contamination}\label{sec:simulations}

The measurement of the number counts requires an estimate of the completeness of the survey, the contamination rate, and the magnitude of the flux boosting, an effect that systematically increases the measured flux densities of sources detected at relatively low SNRs.

\begin{figure}[h]
\includegraphics[width=0.43\textwidth]{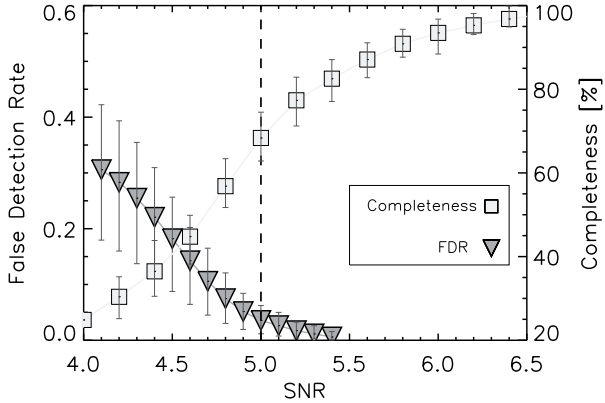}
\caption{Estimated contamination from spurious sources (dark gray triangles) and completeness (light gray squares) and as a function of SNR of the detected sources. The $5\sigma$ adopted threshold is represented by the dashed vertical line, where the false detection rate and completeness are expected to be $\lesssim5\%$ and $\gtrsim70\%$, respectively.  \label{fig:fdr}}
\end{figure}

To estimate the contamination from false detections, we repeat the source extraction procedure described above after inverting all of the 3\,mm continuum maps. All the peaks present in these inverted maps are expected to be noise fluctuations.
The spurious fraction is then estimated as the ratio of the number of negative-to-positive peaks as a function of SNR, where the errors are estimated through a bootstrapping method. As shown in Figure \ref{fig:fdr}, a false detection rate of $\lesssim5\%$ is expected at our adopted threshold of $\ge5\sigma$. This is in agreement with the results of previous ALMA studies, which determined that the rate of false detections falls close to zero at this SNR threshold (e.g. \citealt{Simpson2015a,Fujimoto2016a,Oteo2016a}).

The completeness and flux boosting were quantified using Monte Carlo simulations. Artificial sources are first injected into the flux maps and then they are recovered with the same source extraction procedure used to build the real source catalog. 
A source is considered recovered if it is detected within a synthesized beam of the input random position. After 100 realizations per SNR bin, ranging from 3.0 to 7.0 in steps of 0.1, we determine a completeness of $\gtrsim70\%$ at $\rm SNR\ge5.0$, increasing up to $\gtrsim95\%$ at $\rm SNR\gtrsim6.0$ (Figure \ref{fig:fdr}). The average flux boosting factor due to Eddington bias, measured as the ratio of output-to-input flux density, is found to be $\sim10\%$ at $5.0\sigma$, with the boost factor falling to $\lesssim5\%$ at $\gtrsim6.0\sigma$. Uncertainties in both completeness and flux boosting are estimated as the standard deviation in each bin of SNR and are then propagated in the estimation of the number counts (see \S\ref{sec:num_counts}).

\section{Number Counts} \label{sec:num_counts}
Though no ALMA 3\,mm number counts exist in the literature so far, galaxy number counts have been well studied at shorter wavelengths ($\lambda=850\rm\,\mu m - 1.3.\,mm$)  using  blind ALMA observations (e.g. \citealt{Hatsukade2013a,Carniani2015a,Fujimoto2016a,Oteo2016a,Franco2018a,Hatsukade2018a}). In this paper, we follow the typical method used previously in those works. The contribution of a source with a deboosted flux sensity, $S_i$, to the cumulative number counts are estimated to be:
\begin{equation}
N_i(S_i)=\frac{1-f_{\rm cont}}{\zeta\,A_{\rm eff}(S_i)},
\end{equation}
where $f_{\rm cont}$ is the estimated fraction of contamination at the measured SNR of the source, $\zeta$ is the corresponding completeness, and $A_{\rm eff}(S_i)$ is the largest integrated area sensitive enough to detect sources with $S\ge S_i$ at our adopted threshold (see Figure \ref{fig:area_vs_depth}). As mentioned in \S\ref{sec:simulations}, a low contamination rate ($\lesssim5\%$) is expected given our conservative selection criterion, while the completeness of the survey is found to be  $\approx70-100\%$ (see Figure \ref{fig:fdr}). 
Finally, the  cumulative number counts, $N(>S)$, is estimated by the sum over all of the sources with a flux density higher than $S$.

\begin{figure*}
\centering
\includegraphics[width=0.7\textwidth]{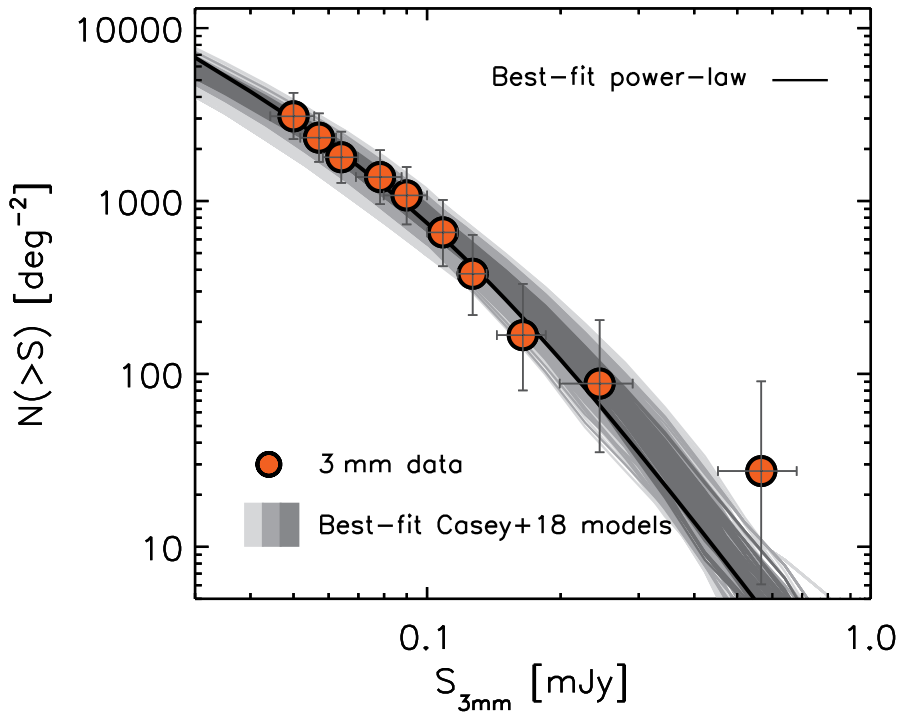}
\caption{Integral galaxy number counts at 3\,mm. The measurements derived in this work are represented by the red points
and the best-fit broken power is plotted as the black solid line. We also plot the number counts from the \citet{Casey2018a,Casey2018b} model when different evolutions on the IRLF are assumed, which are used to fit the data through a maximum likelihood estimation method.
The 68, 95, and 99.7\% confidence intervals for the best-fit models are color-coded from the darkest to the lightest gray, respectively. Their corresponding contributions to the cosmic star-formation rate density are plotted in Figure \ref{fig:sfrd}.
\label{fig:num_counts}}
\end{figure*}

To derive reliable uncertainties in our estimation of the number counts, we take into account the errors associated to the flux densities and  survey's completeness
through a Monte Carlo simulation, where random values are extracted in each realization from  Gaussian distributions with  standard deviations equal to the measured errors. Given our small sample size of 13 sources, Poisson uncertainties are also added in quadrature following \citet{Gehrels1986a}, which are indeed the dominant contributors. 
Figure \ref{fig:num_counts} shows our final cumulative number counts as a function of flux density and the associated uncertainties, after removing the three sources noted in \S\ref{sec:source_extraction} (thought to be either non-thermal or associated with the original targets). 
These estimations are likely to be low biased by cosmic variance since observations across different sightlines were analyzed. 

To parametrize the number counts we  fit a double-power law of the form
\begin{equation}
N(>S)=N_0\left[\left(\frac{S}{S_0}\right)^\alpha + \left(\frac{S}{S_0}\right)^\beta\right]^{-1}, 
\end{equation}
where $N_0$, $S_0$, $\alpha$ and $\beta$ describe the normalization, break, and slope of the power laws, respectively. The best-fit parameters, $N_0=1200^{+1400}_{-1100}$, $S_0=0.11^{+0.22}_{-0.03}\rm\,mJy$, $\alpha=1.4\pm0.5$ and $\beta=3.4\pm0.5$, were inferred using a minimum  $\chi^2$ method through 
a Levenberg-Marquardt algorithm. The resultant best-fit double-power law is plotted in Figure \ref{fig:num_counts}. The number counts were also fitted with a Schechter-like function, but it reproduce neither the behavior of the data at the brightest flux densities nor the shape of the number counts at the faintest end. 

% -> space density at 0.1 mJy = 760 sources/sq. degree
The estimated number counts imply that one serendipitous DSFG is detected at $5\sigma$ per three ALMA Band 3 continuum maps with one hour of integration. This calculation assumes a search area equal to the 1.3 times the FWHM of the primary beam  ($\approx1.3\rm\,arcmin^2$) and a depth equal to $\sigma \approx20\,\mu\rm Jy\,beam^{-1}$). This implies that an even more significant sample of 3\,mm-detected sources can be built using only ALMA archival observations over the next few years.
Similarly, these sources might be detected in the deepest maps achieved at this wavelength with the MUSTANG2 camera on the Green Bank Telescope (e.g. Mroczkowski et al. in preparation), albeit with low angular resolution and  higher integration times compared with ALMA.

\section{Constraining the infrared galaxy luminosity function} \label{sec:models_irlf}

In this section we use the estimated galaxy number counts and the predictions from the backward evolution model presented by \citet{Casey2018a,Casey2018b} to constrain the IRLF, and thus, the contribution of DSFGs to the CSFRD. The model first adopts an infrared galaxy luminosity function of the form
\begin{equation}
\Phi(L,z) =
  \begin{cases} 
      \Phi_\star(z)\left(\frac{L}{L_\star(z)}\right)^{\alpha_{LF}(z)},               & \mbox{if } L<L_\star, \\
      \Phi_\star(z)\left(\frac{L}{L_\star(z)} \right)^{\beta_{LF}(z)},               & \mbox{if } L\ge L_\star,
   \end{cases}
\end{equation}
and assumes an evolution between $0<z\lesssim10$. At $z\lesssim2$, the evolution is well constrained by direct measurements of the IRLF from single-dish telescopes datasets, however, at higher redshifts different evolutions are explored. 
Each galaxy extracted from the assumed IRLF is then assigned an SED according to its luminosity and redshift, following the luminosity-dust temperature (or $L_{\rm IR}-\lambda_{\rm peak}$) relation and correcting for  CMB effects (\citealt{da-Cunha2013a}). Finally, mock observations of the sky are obtained at different wavelengths, areas, and depths, which are used to generate mock measurements of number counts and redshift distributions. The reader is referred to \citet{Casey2018a,Casey2018b} for a thorough description of the model and how sources' flux densities map to the modeled IRLF.

The characteristic number density of the luminosity function, $\Phi_\star$, is assumed to evolve as $(1+z)^{\psi_1}$ 
with a redshift turnover, $z_{\rm turn}\approx2$, from which the relation evolves at higher redshifts with a different slope, $\psi_2$, so that:
\begin{equation}
\Phi_\star \propto \begin{cases}
    (1+z)^{\psi_1}, & \text{if $z<z_\mathrm{turn}$},\\
    (1+z)^{\psi_2}, & \text{if $z>z_\mathrm{turn}$}.
  \end{cases}
\end{equation}
As discussed in the works by \citeauthor{Casey2018a}, the parameters $\psi_1$ and  $z_{\rm turn}$ are fixed to reproduce direct measurements of the luminosity function and the CSFRD at $z\lesssim2.5$, while $\psi_2$ is unconstrained. In this work, we explore different values for this parameter ranging from $\psi_2=-6.5$ to $-2.0$, which map to a very dust-poor early Universe and to an extremely dust-rich one, respectively. Additionally, an extra parameter, $z_{\rm cutoff}$, is used in this analysis to define a redshift above which no more dust-rich galaxies exist, ranging from $z_{\rm cutoff}=9$ down to 5. 
The different evolutionary models of the IRLF are combined with the modeled SEDs to create mock observations from which the number counts are derived. The simulated number counts are then used to fit the measurements derived in \S\ref{sec:num_counts} through a maximum likelihood approach, using flat prior distributions for both $\psi_2$ and $z_{\rm cutoff}$. The corresponding 3\,mm number counts from these models are  shown in Figure \ref{fig:num_counts}, where the best-fit 1, 2, and $3\sigma$ confidence intervals are illustrated by the different colors.

\begin{figure}[h]
\hspace{0.5cm}
\includegraphics[width=0.43\textwidth]{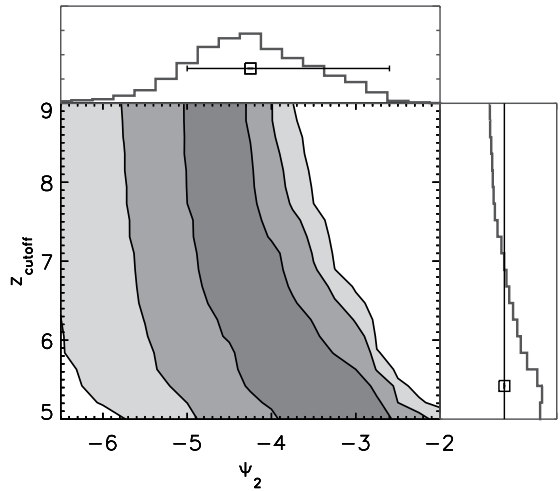}\vspace{0.4cm}
\caption{68, 95, and 99.7\% confidence intervals (represented by the contours in the gray-scale image) for the two parameters explored in this work to describe the evolution of the IRLF in the \citet{Casey2018a,Casey2018b} models. The confidence intervals were determined by fitting the corresponding predicted number counts through a maximum likelihood approach. The derived probability distribution for each individual parameter is represented by the solid line in the external panels (top and right, respectively).
\label{fig:mcmc}}
\end{figure}

\begin{figure}[h]
\includegraphics[width=0.49\textwidth]{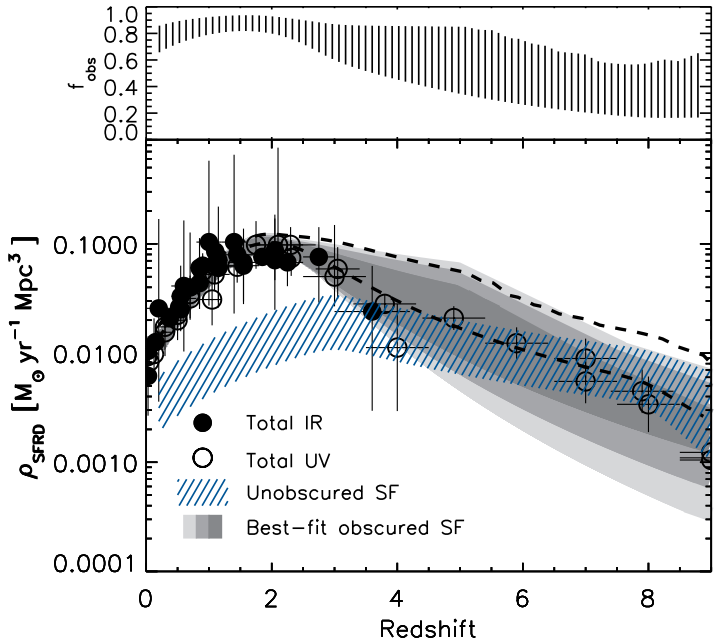}
\caption{ The cosmic star formation rate density as a function of redshift, comparing measurements from rest-frame UV/optical to the obscured component constrained by our 3\,mm number counts. Black circles represent measurements from the literature from both dust-corrected UV (empty circles) and IR rest-frame (filled circles) studies (\citealt{Madau2014a}), which are dominated by UV surveys at $z\gtrsim3.5$. The implied star formation rate densities of the \citet{Casey2018a,Casey2018b}  models that best fit the measured 3\,mm number counts are illustrated  by the gray regions, where the darkest gray represents the 68\% confidence interval (the 95, and 99.7\% are plotted with lighter grays, respectively). On the other hand, the unobscured sources' contribution derived from UV-based measurements, uncorrected by dust attenuation, is represented by the blue hashed region. The implied  total (obscured $+$ unobscured) CSFR is represented by the region delimited by the dashed black lines (68\% confidence interval).  Finally, the fraction of obscured star formation ($\rm SF_{IR}/SF_{UV+IR}$) as a function of redshift derived from the best-fit model predictions is shown in the top panel.
\label{fig:sfrd}}
\end{figure}

As shown in Figure \ref{fig:mcmc}, the best-fit values measured from the 3\,mm number counts provide weak constraints on $z_{\rm cutoff}$ but stronger constraints on $\Psi_2$ with a best-fit value of $\Psi_2=-4.2^{+1.6}_{-0.8}$. Though the range of consistent models is large, indicating significant uncertainty in the yet small 3\,mm sample, we highlight that a dust-poor Universe where DSFGs contribute negligibly ($<10\%$) to the CSFR at $z>4$ ($\psi_2<-5.3$) is not favoured by the data. 
And yet such a sharp downward evolution in the IRLF is  widely assumed in the literature (e.g. \citealt{Finkelstein2015a,Bouwens2015a}).
Actually, as shown in Figure \ref{fig:sfrd}, the best-fit models predict that DSFGs contribute $\approx35-85\%$ of the total CSFRD at $z\approx4-5$ ($68\%$ confidence interval). This implies that the current measurements of the total CSFR at high redshifts, which are based mostly in UV/optical studies of Lyman Break Galaxies samples, might be underestimated up to a factor of $\sim5$. 
At higher redshifts  ($z>5$), due the degeneracy between $z_{\rm cutoff}$ and $\psi_2$ (see Figure \ref{fig:mcmc}), two  scenarios can be constrained. A low  $z_{\rm cutoff}$ value of $\approx6$ implies that the contribution from DSFGs to the CSFRD is indeed negligible at $z\gtrsim6$, but as high as the current measurements derived from surveys tracing the unobscured star formation at $z\lesssim6$. In other words, DSFGs contribute up to $\approx75\%$ of the total star formation rate density up to $z\sim6$. Beyond this redshift, the total CSFRD would  be represented by the current measurements obtained from UV/Optical surveys. On the other hand, if DSFGs are allowed to exist in the model up to $z_{\rm cutoff}\sim9$, the contribution per redshift bin is non-negligible even at $z\sim9$ (although with a large range of uncertainty of $\approx15-65\%$). This later scenario would imply that the current measurements of the total CSFRD are thus biased even at the highest redshifts. 

A further analysis of the assumed dust optical depth is important to understand any possible bias in these estimations. Although, in the model, the SEDs are parametrized as a function of $\lambda_{\rm peak}$ instead of dust temperature, the impact of the CMB on the heating and detection of these sources is a strong function of the dust temperature, and hence, the choice of the dust optical opacity introduces some differences on the galaxies' detected flux  densities at the highest redshifts. The constraints described above assume optically thick SEDs at rest-frame $\lambda<100\,\rm\mu m$ due to the dust self-absorption often present in highly obscured systems. This scenario is supported by the high dust mass typically measured for these galaxies (e.g. \citealt{Michaowski2010a}; \citealt{Magdis2012a}). However, if an optically thin SED is adopted in the model, the  number density of galaxies required to reproduce the same number counts is higher, given the stronger effect of the CMB 
due to the lower dust temperatures associated to the optically thin assumption.
Consequently,  the Universe would need to be more dust-rich than the one predicted by the optically thick assumption discussed above.

Tighter constraints on the obscured CSFRD can be obtained from the redshift distribution of these sources, 
which can be directly compared to the predictions from the model.
Actually, the best-fit models predict that $\approx80-90\%$ of the sources with $S_{3\rm mm}>50\, \mu\rm Jy$ (as those reported here) lie at $z>2$ and  $\approx15-35\%$ at $z>4$ ($68\%$ confidence interval). The high-redshift tail of the expected redshift distribution is then more significant than the ones measured for galaxies selected at shorter wavelengths. For example, \citet{Danielson2017a} found that only $\sim10\%$ of the galaxies selected at $870\,\mu\rm m$ are at $z>4$. However, only 4 of the sources in our catalog have spectroscopic redshifts (see Table \ref{table:catalogue}) and, although very low redshift solutions can be discarded, the current ancillary data is not enough to derive precise photometric redshifts for all the sources since some of them lie outside of the deep imaging surveys. 
The analysis of the redshift distribution of these galaxies will hence be presented in a subsequent work along with recently accepted follow-up ALMA Cycle 6 observations (PI: J. Zavala).

{\twocolumngrid
%Dicuss optically thin model.
\section{Conclusions} \label{sec:conclusions}

We have exploited the ALMA archive to conduct a blind search of serendipitously-detected 3\,mm continuum sources, an extension of the submillimeter-galaxy selection technique to detect Dusty Star-Forming Galaxies at high redshifts. The analyzed data cover a total area of $\approx200\,\rm arcmin^2$, which is equivalent to the area of $\approx240$ ALMA primary beams in this band, an order of magnitude larger than the areas mapped to date in blank-field contiguous observations with ALMA. After masking out the observations' primary targets, we have detected 16 sources above the adopted conservative treshold of $5\sigma$, at which the expected false detection rate is $<5\%$. Using these sources we have derived the first number counts at 3\,mm and estimated that one source is expected per three ALMA Band 3 maps for one hour of integration.

Using the predictions of a backward evolution model, we have found that a dust-poor Universe where DSFGs contribute negligibly to the CSFR at $z>4$, as commonly adopted in the literature, is not favoured by the data. The best-fit models for the evolving IRLF predict that DSFGs contribute $\approx35-85\%$ of the total CSFRD across $z\approx4-5$. At higher redshifts the contribution from dust-obscured star formation is less constrained due to the degeneracy between parameters in the model.
The limits of our constraints themselves could be represented by two broadly different scenarios: A high obscured contribution up to $\approx75\%$ to the total CSFRD up to $z\sim6$, above which the obscured star formation is much more rare, or a non-negligible but more uncertain contribution ($\approx15-65\%$) up to $z\sim9$.
Since these dust-obscured galaxies are  not included in the UV/Optical studies, from which most of the measurements of the CSFRD at high-redshift have been derived, this work suggests that our current understanding of the CSFRD at $z>4$ is still incomplete.

This work highlights the power of 3\,mm observations to detect DSFGs and to measure the dust obscured star formation rate density at the earliest epochs, even when spectroscopic redshifts for individual sources are not available. Given the practice of carrying out millimeter spectral line surveys at this wavelength, a large number of serendipitous detections of 3\,mm continuum sources are expected during the next few years, from which more robust constraints on the dust-obscured star formation rate density  can be derived as well as on  early Universe dust production mechanisms and   galaxy formation and evolution models.

}

\acknowledgments

This work would not have been possible without the rich data available through the ALMA Science Archive.

We thank the reviewer for a helpful report which improved the clarity of the paper.
JAZ and CMC 
thank the University of Texas at Austin College of 
Natural Sciences for support. We also thank NSF grant AST-1714528 and 1814034. EdC gratefully acknowledges the Australian Research Council for funding support as the recipient of a Future Fellowship (FT150100079).
 
This paper makes use of the following ALMA data:
ADS/JAO.ALMA\#2013.1.00092.S, ADS/JAO.ALMA \#2015.1.00853.S, ADS/JAO.ALMA\#2016.1.00171.S, ADS/JAO.ALMA\#2016.1.00567.S, ADS/JAO.ALMA \#2016.1.00932.S, ADS/JAO.ALMA\#2015.1.00228.S, ADS/JAO.ALMA\#2015.1.01151.S, ADS/JAO.ALMA \#2016.1.00324.L, ADS/JAO.ALMA\#2016.1.00754.S, ADS/JAO.ALMA\#2016.1.00967.S, ADS/JAO.ALMA \#2015.1.00752.S, ADS/JAO.ALMA\#2015.1.00862.S, ADS/JAO.ALMA\#2015.1.01222.S, ADS/JAO.ALMA \#2016.1.00698.S, ADS/JAO.ALMA\#2016.1.00798.S, ADS/JAO.ALMA\#2016.1.01546.S, ADS/JAO.ALMA \#2016.1.00564.S, ADS/JAO.ALMA\#2016.1.01149.S. ALMA is a partnership of ESO (representing its member states), NSF (USA) and NINS (Japan), together with NRC (Canada) and NSC and ASIAA (Taiwan) and KASI (Republic of Korea), in cooperation with the Republic of Chile. The Joint ALMA Observatory is operated by ESO, AUI/NRAO and NAOJ.

\facilities{ALMA}

\bibliography{biblio}

\appendix
% \twocolumngrid
\renewcommand\thefigure{\thesection.\arabic{figure}}    

\section{On the possible selection biases}\label{appendix}
\setcounter{figure}{0}

While some of the archival observations used in this work are actually unbiased blank-field observations, the original targets of some other projects might 
introduce some biases on the estimated space density of the 3\,mm-selected galaxies, particularly if the sources targeted are associated with over-dense regions, are known to have a strong clustering, 
or exhibit a high source multiplicity. In this section we investigate the selection biases of each of the 13 sources used in the number counts estimation and conclude that our estimations are not significantly biased.  \\

{\bf ALMA-3mm.02, ALMA-3mm.10:} Both sources were found in the same ALMA observations (\#2015.1.00862.S), which target the $\rm CO(3-2)$ transition in a $z=2.24$ galaxy selected from wide and deep narrow-band  $H_\alpha$ surveys (RA$=$02:16:45.8, Dec$=-$05:02:44.7; \citealt{Sobral2013a,Molina2017a}). The original source is not detected in the ALMA continuum image, from which we derive a flux density upper limit of $S_{\rm 3mm}<21\rm\,mJy$ ($3\sigma$). This a factor of 6 and 3 fainter than our 3\,mm candidates ALMA-3mm.02 and ALMA-3mm.10, which are located at 31 and 28 arcsec from the main target, respectively. All this information suggests that our source candidates are not related to the original one. Actually,  ALMA-3mm.10 is part of an independent ALMA follow-up of AzTEC sources at 1.1\,mm (\citealt{Ikarashi2017a}), and despite having a similar flux density to other sources in the catalog whose redshift distribution peaks between $z=2-3$, it lacks a photometric redshift estimation, suggesting a higher redshift solution.

{\bf ALMA-3mm.03:} This source was found in the observations of the ASPECS project  (ALMA project code: \#2016.1.00324; see also \citealt{Walter2016a}), which by design is an ubiased blank-field survey. 

{\bf ALMA-3mm.04:} The source was detected in the ALMA program \#2016.1.00698.S, which aims to detect the Sunyaev--Zeldovich effect in a galaxy cluster at $z=1.91^{+0.19}_{-0.21}$  (\citealt{Mantz2014a}). A further analysis of the extracted spectrum of this candidate revealed a line at 105.27\,GHz (Zavala et al. in preparation), which is inconsistent with the cluster redshift (the closest solution  to the cluster's redshift is $z=2.28$). Therefore, this indicates that our galaxy is not associated with the cluster structure.
% , although some gravitational amplification might be present. 
% If we adopt the redshift solution to be the one which is closest to the cluster's redshift, we derive an spectroscopic redshift of $z=2.28$, indicating that our galaxy is not associated with the cluster structure, although some gravitational amplification might be presented. 

{\bf ALMA-3mm.06, ALMA-3mm.13:} The ALMA observations (\#2015.1.01151.S) target the $\rm CO(2-1)$ transition in galaxies within a proto-cluster at $z=1.6$. Despite being detected in the continuum maps, our detections do not show any features at $\sim 88.2$\,GHz, the expected frequency for the $\rm CO(2-1)$ line given the proto-cluster redshift. Indeed, one of our sources (ALMA-3mm.13), shows a line at $\sim 99.8$\,GHz (Zavala et al. in preparation), which confirms that the source is not part of the targeted structure. The only galaxy found in the $z=1.6$ proto-cluster (\citealt{Noble2017a}; RA$=$03:30:59, Dec$=-$28:43:06) is actually not detected in the 3\,mm continuum map.

{\bf ALMA-3mm.07, ALMA-3mm.09, ALMA-3mm.14:} These three objects were found in the same project (\#2016.1.00754.S), which comprises spectroscopic
observations towards multiple submillimeter galaxies selected based on previous $870\,\rm\mu m$ ALMA continuum imaging. Given the nature of the targeted galaxies (multiple dusty star-forming galaxies), 
our 3\,mm galaxies might be related to the $870\,\rm\mu m$-selected multiple galaxies, introducing bias and breaking the  blindness of our selection. However, here we show that these sources are most likely not related to the main sample and hence they can be considered blind detections. ALMA-3mm.07 (ALESS75.C in \citealt{Wardlow2018a}), was serendipitously detected in the same field as the $870\,\rm\mu m$-detected galaxies ALESS75.1 and ALESS75.2 (the original targets of the pointing), which lie at $z_{\rm spec}=2.545$ and 2.294, respectively (\citealt{Danielson2017a}). 
Our source does not show any emission line at the expected frequency of the  $\rm CO(2-1)$ line at $z=2.5$. Unfortunately, the ALMA tunings do not cover the frequencies of the expected CO lines for the $z=2.29$ solution. Nevertheless, this source shows a very different color ratio from the program's sources
($S_{870\rm\mu m}/S_{3\rm mm}<17$ vs 114  and $>714$, respectively), and is even not detected at $870\,\rm\mu m$.  
Actually, \citealt{Wardlow2018a} reported that this source, ALMA-3mm.07, has a photometric redshift of $4.00^{+0.07}_{-0.08}$ (\citealt{Taylor2009a,Cardamone2010a}), which is inconsistent with the redshifts of the two original targeted sources. 
Similarly,  ALMA-3mm.09 (ALESS87.C in \citealt{Wardlow2018a}) shows no emission lines at the searched frequencies and is not detected at $870\,\rm\mu m$, contrary to the program's targets (ALESS87.1 and ALESS87.2). Its $S_{870\rm\mu m}/S_{3\rm mm}$ ratio is also very different from the project's sample ($<22$ vs 28 and $>54$, respectively) and, as mentioned by \citet{Wardlow2018a}, is in better agreement with a  $z>4$ galaxy.  Finally, ALMA-3mm.14 is more than 30 arcsec away from the original $870\,\rm\mu m$-selected objects, and consequently is outside of the ALMA primary beam at that frequency. This source does not show any emission line either, suggesting again a very different redshift solution. 

{\bf ALMA-3mm.08, ALMA-3mm.15, ALMA-3mm.16:} The three sources were detected in different observations of the same project (\#2016.1.00171.S), whose targets are galaxies at $z\sim1.1$. As described in detail in the main text, our galaxies are expected to lie at $z\gtrsim2.5$, and therefore our detections can be considered unbiased.  In fact, ALMA-3mm.08 has a spectroscopic redshift of $z_{\rm spec}=4.55$ (see Table \ref{table:catalogue}), significantly higher than the original sample. Furthermore, ALMA-3mm.15 and ALMA-3mm.16 have been found $\gtrsim30$ arcsec away from the center of the maps, where the main targeted sources are located. This further confirms that those galaxies are not associated with the program's sample.

{\bf ALMA-3mm.12:} The project where this source was found targets a sample of  starburst galaxies at $z\sim1.6$ (\#2015.1.00861.S), nevertheless, the 3\,mm galaxy has a spectroscopic redshift of $z_{\rm spec}=2.99$ (see Table \ref{table:catalogue}), indicating that the original sample selection does not introduce any particular bias in our detection.\\

\begin{figure}[ht]
\begin{center}
\includegraphics[width=0.45\textwidth]{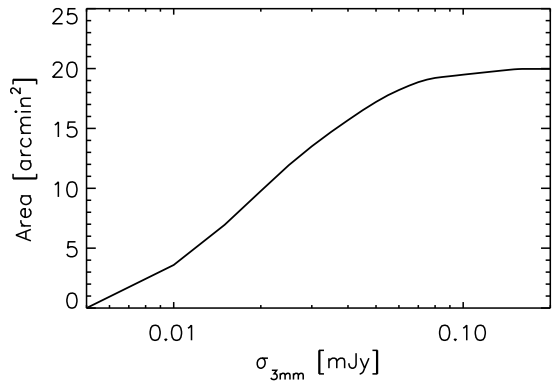}
\caption{Cumulative area covered by those ALMA maps with blind detections as a function of $1\sigma$ r.m.s. As it can be seen, the depths probed by these maps encompass a large range from $\sim 6$ to $100\rm\,\mu Jy$.
\label{fig:area_vs_depth_appendix}}
\end{center}
\end{figure}

As it has been shown, the 3\,mm detected galaxies used in the estimation of the number counts are most likely not related to the original targets of the observations. Actually, most of the original project sources are not detected in the 3\,mm continuum images, and the 3\,mm-selected galaxies show properties very different from the original targeted objects. This confirms the uniqueness of our selection criteria which likely selects dusty star-forming galaxies at high redshifts. The only possible bias which might be present in our analysis is the gravitational lensing associated with those observations towards clusters of galaxies. Nevertheless, the redshifts of these clusters ($z=1.6$ and 1.9, respectively) are significantly higher than the typical lens clusters ($z<1$; e.g. \citealt{Zavala2015a}) and, furthermore, the probability of amplification depends not only on the sources' redshift but also on the angular offset from the cluster position, making unlikely the presence of strong lensing. Besides, these observations only represent $\sim6\%$ of the total analyzed area. On the other hand, the amplification by foreground large-scale structures has also been found in blank-field observations (e.g. \citealt{Aretxaga2011a}). All this analysis indicates that our detections are low biased by the original sample selection of the observations and, if any bias is present, it is similar to the one that can plague any other blind survey.
Additionally, the maps in which these sources were detected have a large range of depths representative of the whole survey  (Figure \ref{fig:area_vs_depth_appendix}). 
Finally, we highlight that the rest of the used archival ALMA observations in which no sources were detected show a similar heterogeneous sample selection and depths, and therefore, our whole survey is not expected to be significantly biased.

\end{document}